\documentclass[aps,prl,twocolumn,superscriptaddress]{revtex4}
\usepackage{graphicx}
\usepackage{here}
\begin{document}
\title{Mechanism of the close-to-threshold production of the $\eta$ meson}
\author{R.~Czy{\.z}ykiewicz}
\author{P.~Moskal}
\affiliation{Institute of Physics, Jagellonian University, 30-059 Cracow, Poland}
\affiliation{IKP \& ZEL, Forschungszentrum J\"ulich, 52425 J\"ulich, Germany}
\author{H.-H.~Adam}
\affiliation{IKP, Westf\"alische Wilhelms-Universit\"at, 48149 M\"unster, Germany}
\author{A.~Budzanowski}
\affiliation{Institute of Nuclear Physics, 31-342 Cracow, Poland}
\author{E.~Czerwi\'nski}
\author{D.~Gil}
\affiliation{Institute of Physics, Jagellonian University, 30-059 Cracow, Poland}
\author{D.~Grzonka}
\affiliation{IKP \& ZEL, Forschungszentrum J\"ulich, 52425 J\"ulich, Germany}
\author{M.~Hodana}
\affiliation{Institute of Physics, Jagellonian University, 30-059 Cracow, Poland}
\author{M.~Janusz}
\affiliation{Institute of Physics, Jagellonian University, 30-059 Cracow, Poland}
\affiliation{IKP \& ZEL, Forschungszentrum J\"ulich, 52425 J\"ulich, Germany}
\author{L.~Jarczyk}
\author{B.~Kamys}
\affiliation{Institute of Physics, Jagellonian University, 30-059 Cracow, Poland}
\author{A.~Khoukaz}
\affiliation{IKP, Westf\"alische Wilhelms-Universit\"at, 48149 M\"unster, Germany}
\author{K.~Kilian}
\affiliation{IKP \& ZEL, Forschungszentrum J\"ulich, 52425 J\"ulich, Germany}
\author{P.~Klaja}
\affiliation{Institute of Physics, Jagellonian University, 30-059 Cracow, Poland}
\affiliation{IKP \& ZEL, Forschungszentrum J\"ulich, 52425 J\"ulich, Germany}
\author{B.~Lorentz}
\author{W.~Oelert}
\affiliation{IKP \& ZEL, Forschungszentrum J\"ulich, 52425 J\"ulich, Germany}
\author{C.~Piskor-Ignatowicz}
\affiliation{Institute of Physics, Jagellonian University, 30-059 Cracow, Poland}
\author{J.~Przerwa}
\affiliation{Institute of Physics, Jagellonian University, 30-059 Cracow, Poland}
\affiliation{IKP \& ZEL, Forschungszentrum J\"ulich, 52425 J\"ulich, Germany}
\author{B.~Rejdych}
\affiliation{Institute of Physics, Jagellonian University, 30-059 Cracow, Poland}
\author{J.~Ritman}
\affiliation{IKP \& ZEL, Forschungszentrum J\"ulich, 52425 J\"ulich, Germany}
\author{T.~Sefzick}
\affiliation{IKP \& ZEL, Forschungszentrum J\"ulich, 52425 J\"ulich, Germany}
\author{M.~Siemaszko}
\affiliation{Institute of Physics, University of Silesia, Katowice, Poland}
\author{J.~Smyrski}
\affiliation{Institute of Physics, Jagellonian University, 30-059 Cracow, Poland}
\author{A.~T\"aschner}
\affiliation{IKP, Westf\"alische Wilhelms-Universit\"at, 48149 M\"unster, Germany}
\author{K.~Ulbrich}
\affiliation{ISK, Rheinische Friedrich-Wilhelms-Universit\"at, 53115 Bonn, Germany}
\author{P.~Winter}
\affiliation{Department of Physics, University of Illinois at
Urbana-Champaign, Urbana, IL 61801, USA}
\author{M.~Wolke}
\affiliation{IKP \& ZEL, Forschungszentrum J\"ulich, 52425 J\"ulich, Germany}
\author{P.~W\"ustner}
\affiliation{IKP \& ZEL, Forschungszentrum J\"ulich, 52425 J\"ulich, Germany}
\author{W.~Zipper}
\affiliation{Institute of Physics, University of Silesia, Katowice, Poland}
\date{\today}
\begin{abstract}
Measurements of the analysing power for the $\vec{p}p\to pp\eta$
reaction have been performed 
in the close-to-threshold energy region 
at beam momenta of p$_{beam}$~=~2.010 and
2.085~GeV/c, corresponding to excess energies of Q~=~10 and 36~MeV, 
respectively. 
The determined analysing power is essentially consistent with zero 
implying that the $\eta$ meson is produced predominantly in the $s$-wave
at both excess energies. 
The angular dependence of the  analysing power,
combined with the hitherto determined 
isospin dependence of the total cross section for the $\eta$ meson production in 
nucleon-nucleon collisions, reveal a statistically significant 
indication that the excitation of the nucleon to the 
S$_{11}$(1535) resonance, the process which intermediates the 
production of the $\eta$ meson,  
is predominantly due to the exchange of the 
$\pi$ meson between the colliding nucleons.  
\end{abstract}
\maketitle

In the low energy region investigations of the production
and decay of hadrons 
deliver information needed to deepen our knowledge about
the strong interaction in the domain where the perturbative approach is not possible.

In this letter we focus on the hadronic production mechanism
of the $\eta$ meson 
and interpret 
the empirical observations
in the framework of meson exchange models. 
We report on the determination of the 
angular dependence of the beam
analysing power for the $\vec{p}p\to pp\eta$ reaction.  
We also demonstrate that 
the confrontation of predictions based upon different
scenarios, involving exchanges of various mesons,  with 
the so far determined unpolarised observables and with 
results on the analysing power,
permits to single out the dominant process and hence to understand  
details of the $\eta$ meson production %in the collsions of nucleons
close to the kinematical threshold. 

\begin{figure}[h]
\parbox{0.11\textwidth}{
\includegraphics[width=0.14\textwidth]{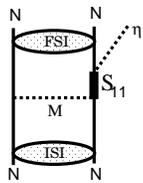}}
\parbox{0.36\textwidth}{\caption{
The mechanism of the $\eta$ meson
production in nucleon-nucleon collisions: excitation of a
nucleon to the S$_{11}$(1535) resonance state via meson
exchange, and its subsequent decay into the
proton-$\eta$ system. M denotes an intermediate pseudoscalar or
vector meson, e.g. $\pi$, $\eta$, $\omega$, $\rho$.
ISI and FSI indicate initial and final state interaction
between the nucleons.
\label{pion}
}}
\end{figure}

From precise measurements of the total cross sections
of the $\eta$ meson production in the $pp\to pp\eta$
reaction~\cite{hibou,jureketa,bergdolt,chiavassa,calen1,calen2,moskal-prc,abdel}
it was concluded~\cite{moalem,bati,germond,laget,vetter,oset,nakayama,wilkin} 
that this process proceeds through the
excitation of one of the protons to the S$_{11}$(1535)
state which subsequently deexcites via the emission of the $\eta$ meson
(see Fig.~\ref{pion}).

\vspace{-0.6cm}
\vspace{-0.3cm}
The crucial observations were a large value of the absolute
cross section and an isotropic distributions
of the emission angle of the $\eta$ meson in the reaction
center-of-mass system. In practice, within the meson exchange
picture, the excitation of the intermediate resonance can be
induced by exchange of any of the
pseudoscalar or vector ground state mesons between the nucleons. Based only
on the excitation function it was, however, impossible  
to disentangle the contributions 
to the production process originating from the $\pi$, $\eta$, 
$\omega$ or $\rho$ meson exchange. 
In fact, due to the negligible variation of the production
 amplitude in the range of a few tens of MeV, the only information
 available from the excitation function is reduced to a single 
 number~\cite{bernard259,review,hab}. \\
More constraints to theoretical 
models~\cite{germond,laget,wilkin,moalem,vetter,nakayama,oset,bati}
have been deduced from the measurement
of the isospin dependence of the total cross section
by the WASA/PROMICE collaboration~\cite{calen3}~\footnote{The
measurement of the isospin dependence is being extended by the COSY-11
collaboration~\cite{jpg06} to the $\eta^{\prime}$ production which may also be sensitive
to gluonic production mechanism~\cite{steven}.}
The comparison of the $\eta$ meson production in
proton-proton and proton-neutron collisions
inferred that the $\eta$ meson
is by a factor of twelve more copiously produced when the total
isospin of the nucleons is zero with respect to the case when it 
is one. As a consequence only an isovector meson exchange
is conceivable as being responsible for such a strong
isospin dependence. 
This result was already a large step forward but still
the relative contributions of the $\rho$ and $\pi$ mesons
remained to be disentangled.
The elucidation of this very detail in the production mechanism
of the  $\eta$ meson 
constitutes the motivation for the measurements presented 
in this letter. For this purpose we have determined the  
analysing power for the $\vec{p}p\to pp\eta$ reaction since its 
theoretical value~\cite{wilkin,nakayama} 
is sensitive to the assumption on the type
of the meson being exchanged in order to excite one of the
colliding nucleons to the S$_{11}$(1535) state.\\ 
For the measurement of the $\vec{p}p\to pp\eta$ reaction the
COSY-11 experimental setup~\cite{brauksiepe} has been 
used, which is an internal beam facility at the cooler synchrotron
COSY~\cite{meier,prasuhn}. 
This facility has been described in various
publications, therefore 
herein we only briefly report the main 
idea of the measurement 
referring 
the interested reader
to the articles~\cite{brauksiepe,smyrski} for the description 
of the instrumentation system and to~\cite{klaja,nim,hab} for the
particle identification and monitoring 
of the beam and target parameters.\\  
A vertically polarised proton 
beam~\cite{stockhorst},
had been stored and
accelerated in the COSY ring.
The direction of the polarisation was being flipped
from cycle to cycle.
The target is installed in front of a machine 
dipole magnet acting as a momentum separator for the 
charged reaction products. 
The positively charged ejectiles were registered in 
drift chambers and scintillator hodoscopes. 
For each particle
its trajectory and time of flight on a nine meter distance
was measured.
Tracking back these trajectories through the known 
magnetic field inside the dipole magnet to the reaction 
vertex allows for the momentum reconstruction 
with a precision of about
4~MeV/c~\cite{moskal-prc,hab} in the center-of-mass frame. 
The acceptance of the COSY-11
facility allows to register only events scattered
near the horizontal plane. 
In the analysis the azimuthal angle $\phi$ was restricted 
to values of $\cos\phi$ ranging between 0.87 and 1.\\
For the determination of the analysing power %for the production 
of the $\eta$ meson at a given value of the polar angle $\theta_{\eta}$ and 
the azimuthal angle $\phi$
it is required
to measure a left-right asymmetry of yields of the $\eta$ meson 
in the frame turned by the angle $\phi$ with respect 
to the laboratory coordinate system. This is often referred to as a 
Madison frame~\footnote{
        Madison Convention, 
        {\it Polarisation Phenomena in Nuclear Reactions}, 
        University of Wisconsin Press, Madison, pp. XXV (1971)
} 
which in our case has its $y$ axis parallel to the
$\vec{p}_{beam}\times \vec{p}_{\eta}$ vector,
with $\vec{p}_{beam}$ and $\vec{p}_{\eta}$ denoting the momentum vectors
of the proton beam and the $\eta$ meson 
in the center-of-mass system, respectively. \\ 
In the case of the $pp\to pp\eta$ reaction,
the COSY-11 detector setup, 
has much larger acceptance for 
events where the $\eta$ meson is produced to the
left side with respect to the polarisation plane
as compared to those for which it is emitted to the right. 
\begin{table}[b]
\vspace{-1.2cm}
\caption{       Number of reconstructed $\eta$ mesons 
                and 
          analysing powers for the $\vec{p}p\to pp\eta$ reaction. 
		\label{tab1}}
\begin{ruledtabular}
\begin{tabular}{clllr}
Q [MeV] & $\cos\theta_{\eta}$ & $N^{\uparrow}_{+}$ & $N^{\downarrow}_{-}$ & $A_y$ \\
\hline
           & $\left(-1;-0.5\right)$ & 306$\pm$27 & 250$\pm$26 & {\bf .163}$\pm${\bf .099}$\pm${\bf .022} \\
  {\bf 10} & $\left(-0.5;0\right)$  & 267$\pm$22 & 260$\pm$24 & {\bf .035}$\pm${\bf .091}$\pm${\bf .012} \\
           & $\left(0;0.5\right)$   & 198$\pm$18 & 208$\pm$19 & {\bf -.021}$\pm${\bf .095}$\pm${\bf .011}\\
           & $\left(0.5;1\right)$   & 279$\pm$23 & 286$\pm$25 & {\bf -.003}$\pm${\bf .088}$\pm${\bf .009}\\
    \hline
           & $\left(-0.5;0\right)$  & 103$\pm$16 & 100$\pm$18 & {\bf .039}$\pm${\bf .179}$\pm${\bf .012} \\ 
  {\bf 36} & $\left(0;0.5\right)$   & 144$\pm$16 & 153$\pm$18 & {\bf -.029}$\pm${\bf .122}$\pm${\bf .010}\\
           & $\left(0.5;1\right)$   & 259$\pm$24 & 296$\pm$28 & {\bf -.084}$\pm${\bf .100}$\pm${\bf .011} \\
\end{tabular}
\end{ruledtabular}
\end{table}    
Therefore, the left-right asymmetries
are determined from numbers 
of events with the $\eta$ meson production
to the left side measured for the spin up  and spin down mode 
of the beam polarisation. 
For the quantitative evaluation 
we define $N^{\uparrow}_{+}(\theta_{\eta})$ and
$N^{\downarrow}_{-}(\theta_{\eta})$  
as  production yields of the $\eta$ meson
emitted to the left around the $\theta_{\eta}$
angle as measured with  the up and  down beam polarisation, respectively, i.e.  
\vspace{-0.2cm}
\begin{eqnarray}
N^{\uparrow}_{+}(\theta_{\eta}) = \sigma_{0}\left(\theta_{\eta}\right)\left(1+P^{\uparrow}A_y\left(\theta_{\eta}\right)\right) E(\theta_{\eta}) \int{L^{\uparrow}dt} ,
\label{yields1}
\\
N^{\downarrow}_{-}(\theta_{\eta}) = \sigma_{0}\left(\theta_{\eta}\right)\left(1-P^{\downarrow}A_y\left(\theta_{\eta}\right)\right) E(\theta_{\eta}) \int{L^{\downarrow}dt},
\label{yields2}
\end{eqnarray}
with $\sigma_{0}\left(\theta_{\eta}\right)$ denoting the
cross section for the $\eta$ meson production for 
unpolarised beam, $P^{\uparrow\left(\downarrow\right)}$
standing for the polarisation degree corresponding to spin up and down modes,
E$(\theta_{\eta})$ being the efficiency of the
COSY-11 facility for detecting the $\eta$ meson
emitted to the left side at the $\theta_{\eta}$ angle
and $L^{\uparrow\left(\downarrow\right)}$ denoting the luminosity
during the beam polarisation up and down.
Signs in the brackets of Eqs.~\ref{yields1} and~\ref{yields2}
follow the Madison convention.
\vspace{-0.4cm}
\begin{figure}[h]
\includegraphics[width=4.27cm]{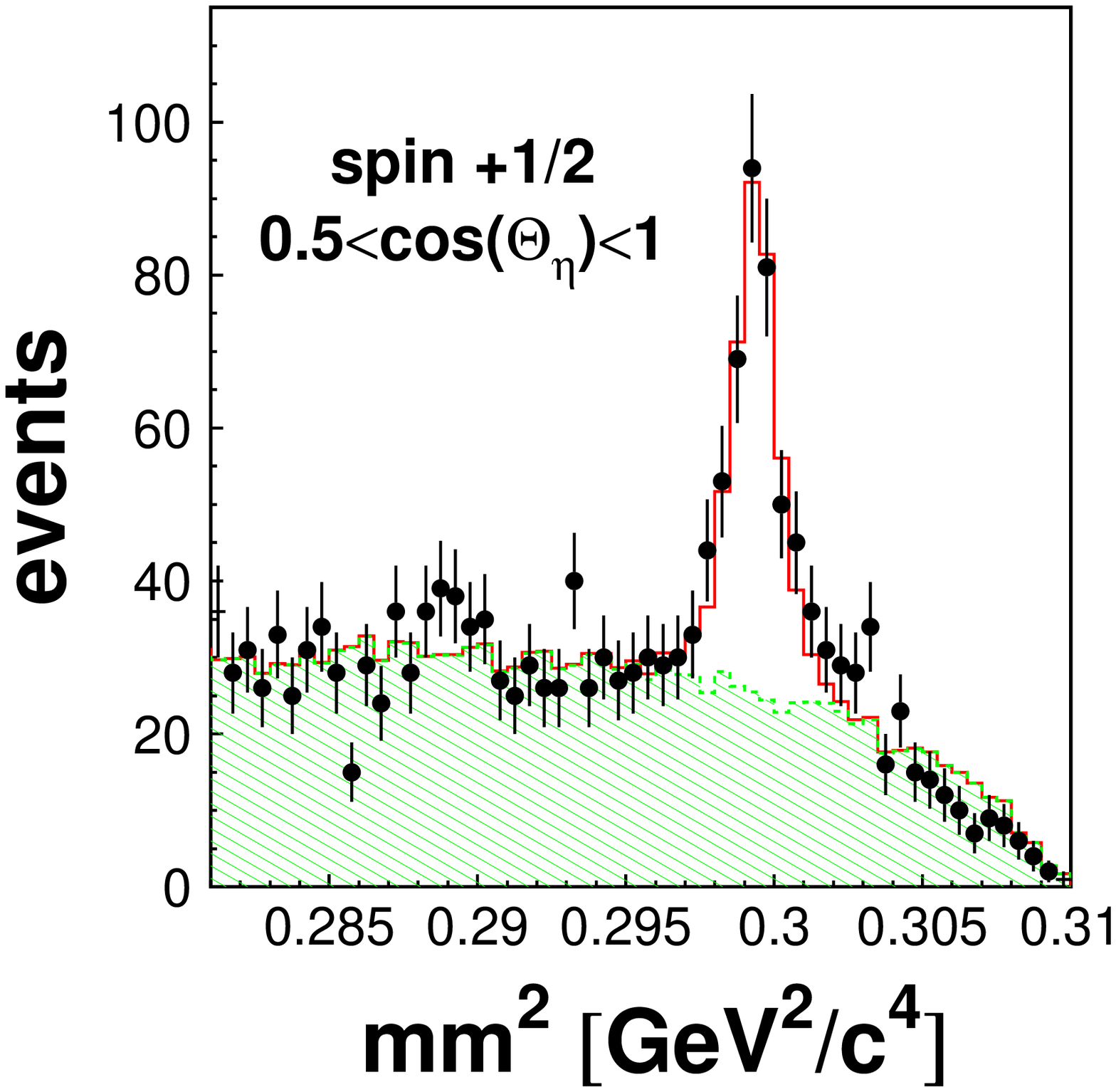}
\includegraphics[width=4.27cm]{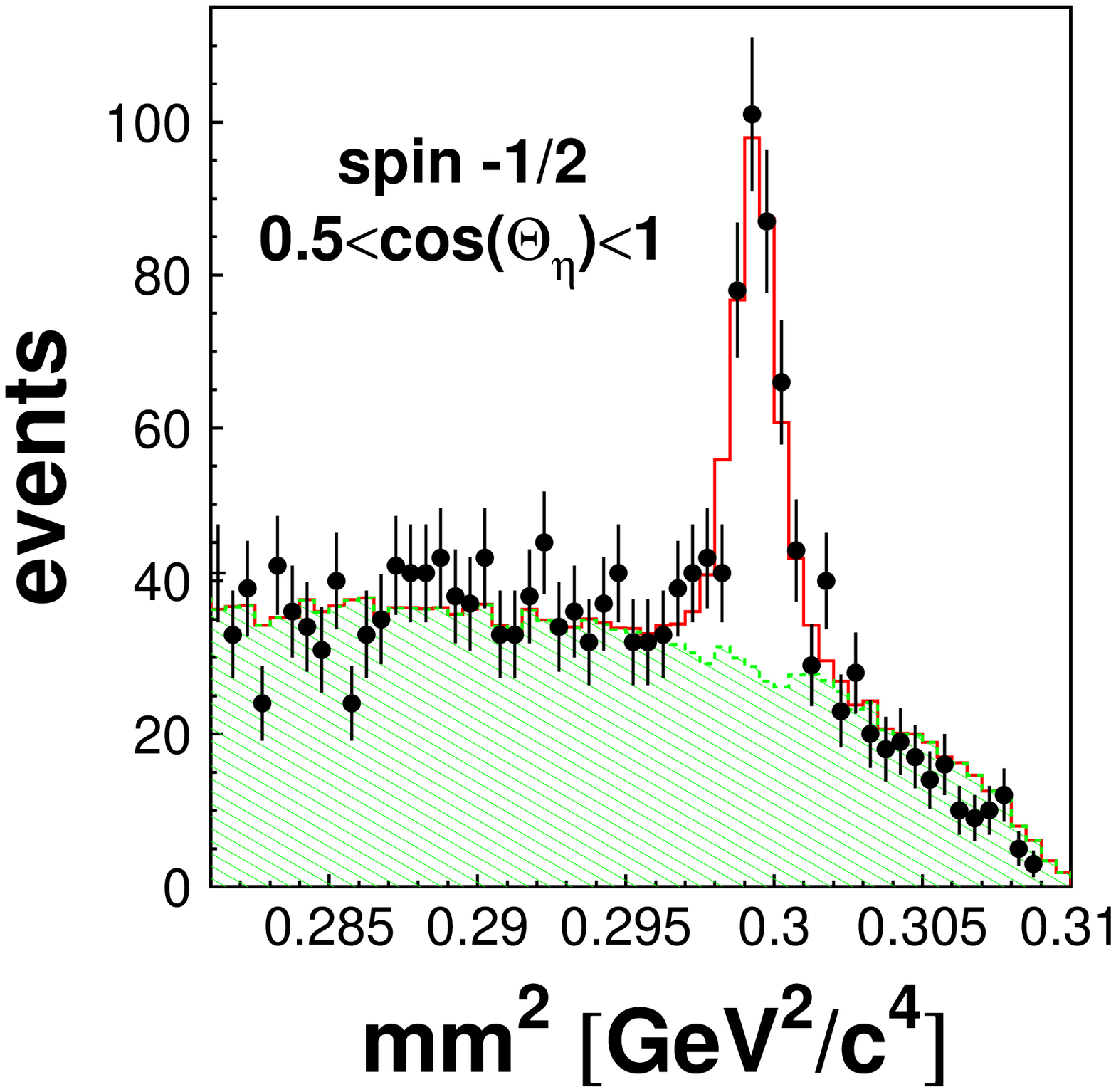}

\vspace{-0.3cm}
\caption{Examples of missing mass spectra for $\cos\theta_{\eta}\in\left(0.5;1\right)$
and opposite beam polarisation states, as measured 
at the excess energy Q~=~10~MeV. 
Full points correspond to the experimental values which are shown with their statistical 
uncertainties. The solid line represents the sum of the $pp\eta$ and multipion
background production channels determined by Monte-Carlo simulations.
The shaded parts of the histograms show the simulated 
contributions from  
the multipion background. \label{miss_mass}}
\end{figure}
\vspace{-0.3cm}
Assuming that $P^{\uparrow}~\approx~P^{\downarrow}$~\footnote{Which is
valid within $\pm$2\% accuracy, as has been studied with the
EDDA~\cite{altmeier} and COSY~\cite{bauer} polarimeters.} 
and introducing the average beam polarisation
$P\approx\frac{P^{\uparrow}+P^{\downarrow}}{2}$,
the relative luminosity
$L_{rel}=\frac{\int{L^{\uparrow}dt}}{\int{L^{\downarrow}dt}}$
and solving Eqs.~\ref{yields1} and~\ref{yields2} for A$_{y}(\theta_{\eta})$
we obtain:
\vspace{-0.3cm}
\begin{equation}
A_y(\theta_{\eta}) = \frac{1}{P} \frac{N^{\uparrow}_{+}(\theta_{\eta})-L_{rel}N^{\downarrow}_{-}(\theta_{\eta})}{N^{\uparrow}_{+}(\theta_{\eta})+L_{rel}N^{\downarrow}_{-}(\theta_{\eta})}.
\label{ayform}
\end{equation}
\vspace{-0.2cm}

The production yields $N^{\uparrow}_{+}(\theta_{\eta})$ and $N^{\downarrow}_{-}(\theta_{\eta})$
have been extracted from the missing mass spectra.
Optimizing the statistics and the expected shape of the 
analysing power function, the range of the  
$\theta_{\eta}$ angle has been divided 
into four bins, at both excess energies. 
Fig.~\ref{miss_mass} presents missing mass spectra
obtained for the measurements at Q~=~10~MeV for $\cos\theta_{\eta}\in\left(0.5;1\right)$
corresponding to different states of the beam polarisation.  
To separate the actual production rates from the background both the 
reactions with multipion production as well as the events 
with the $\eta$ meson production have been simulated using 
a program based on the GEANT3~\cite{geant} code. 
Since we consider here only the very edge of the phase space distributions, i.e.
where the protons are emitted predominantly in the S-wave,
the shape of the background can be reproduced assuming that the homogenous
phase space   distribution is modified only by the interaction between protons~\cite{hab,jpg06}. 
A fit of the 
simulated missing mass spectra to the corresponding experimental
histograms has been performed  
with the amplitudes of the simulated spectra
treated  as free parameters.
The extracted $N^{\uparrow}_{+}(\theta_{\eta})$ and $N^{\downarrow}_{-}(\theta_{\eta})$
values, are quoted in Table~\ref{tab1}
along with their statistical uncertainties. 

The relative luminosity for both excess energies has been determined by means of the 
measurement of  coincidence rate
in the polarisation plane~\cite{czyzyk-phd}.
Due to the parity invariance, 
in the polarisation  plane the differential cross section for any nuclear 
reaction caused by the strong interaction 
is independent of the beam polarisation and therefore 
a ratio of the numbers of events during spin up and down 
modes can be used as a measure of the relative luminosity.
Values of $L^{10}_{rel}=0.98468\pm 0.00056\pm 0.00985$
and $L^{36}_{rel}=0.98301\pm 0.00057\pm 0.00985$
have been obtained at the excess energies of Q~=~10 and 36~MeV, respectively.  

During the run at the excess energy of Q~=~10~MeV the beam polarisation 
has been determined with the COSY-11 setup~\cite{czyzyk1}.
The principle of measurement was based on the
determination of the asymmetry in the accelerator plane (perpendicular to the polarization vector)
for the $\vec{p}p\to pp$ reaction.  
Although at a given beam polarisation mode only 
protons elastically scattered to the right could be registered
it was possible to determine the polarisation by flipping the spin
and employing Eq.~\ref{ayform} with exchanged $A_y\left(\theta_{\eta}\right)$
and $P$.
As a result the value of the polarisation 
degree P~=~0.680~$\pm$~0.007~$\pm$~0.055 has been extracted. 
For the calculations the values of analysing powers 
for the elastic processes have been taken into account from
the precise measurements performed 
by the EDDA collaboration~\cite{altmeier}.
 
For the polarisation monitoring during the run at Q~=~36~MeV
we used the EDDA facility~\cite{altmeier}.
The determination of the polarisation degree with this setup 
is based on the asymmetry measurement for 
the $\vec{p}p\to pp$ process. The obtained value 
of the averaged polarisation 
equals to P~=~0.663~$\pm$~0.003~$\pm$~0.008. 

The main source of the systematic uncertainties 
in the determination of the production yields 
originates from a background misidentification.
In order to estimate the systematic error, 
an alternative method (with respect to the method 
presented above) of background subtraction has been 
applied which is based on a polynomial background cut~\cite{czyzyk-phd}. 
Differences in the production yields obtained by 
this independent method were 
less than 1.5\%. 
The main contribution to the systematic uncertainty 
of the relative luminosity might be due to a slight shift
of the center of detectors outside the polarisation plane. 
Assuming very conservatively 
a 4~mm shift 
and using the analysing powers
for the elastically scattered protons of Ref.~\cite{altmeier} 
a value of 1\% systematic uncertainty of the relative 
luminosity was established by means of 
Monte-Carlo simulations.
The systematic uncertainty of 8\% for the polarisation
measured with the COSY-11 polarimeter
is determined by error propagation from 
Eq.~\ref{ayform} with the systematic uncertainties of the 
analysing powers (1.2\%), systematic error of the relative
luminosity (1\%) and the 1\% systematic uncertainty
of the number of elastically scattered yields~\cite{nim}.  
During the measurement at the excess energy of Q~=~36~MeV 
the overall systematic error 
in the determination of the polarisation value, when using the 
large acceptance EDDA detector, was 1.2\%~\cite{altmeier}.  
\vspace{-0.2cm}
\begin{figure}[h]
\includegraphics[width=4.27cm]{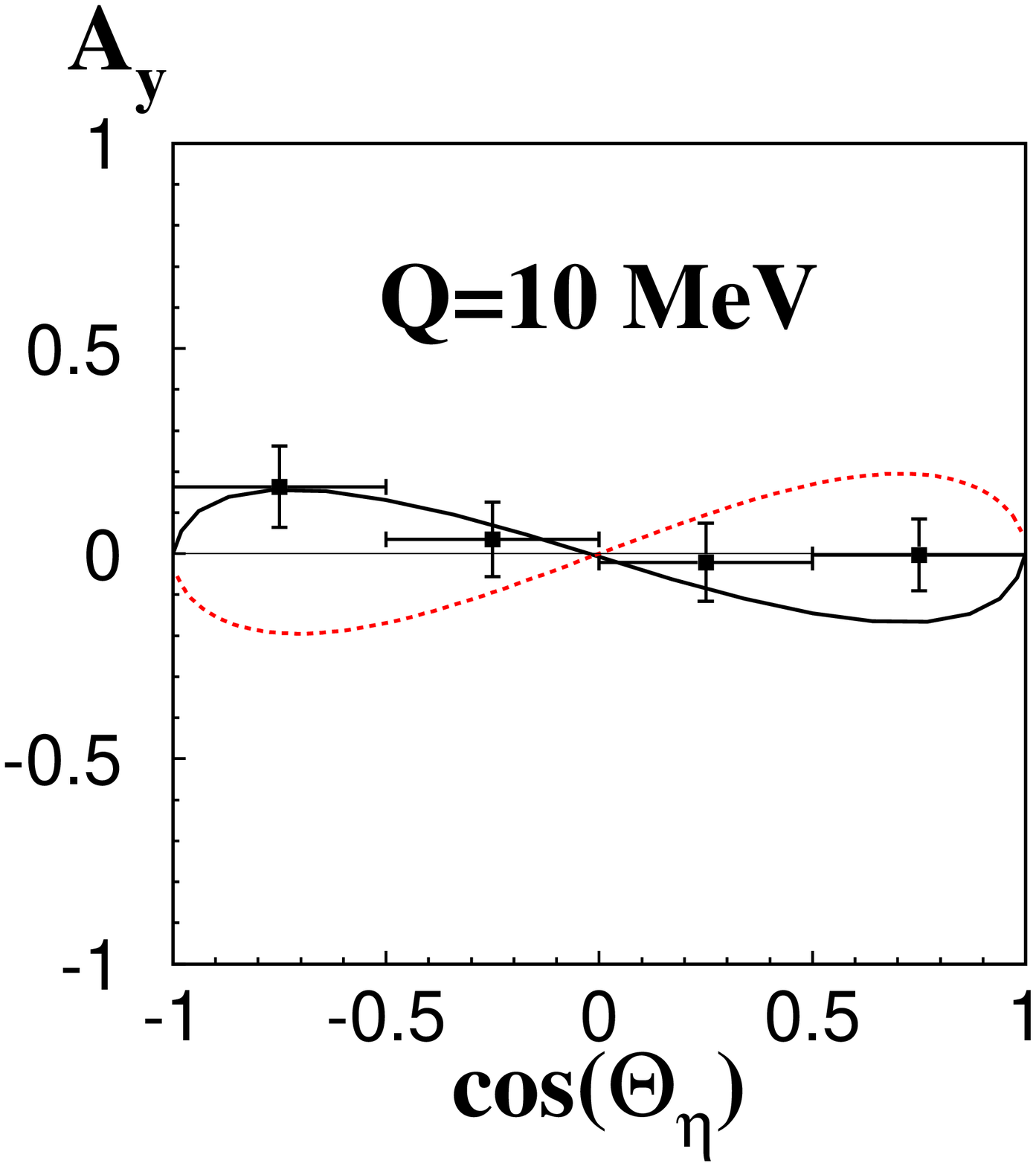}
\includegraphics[width=4.27cm]{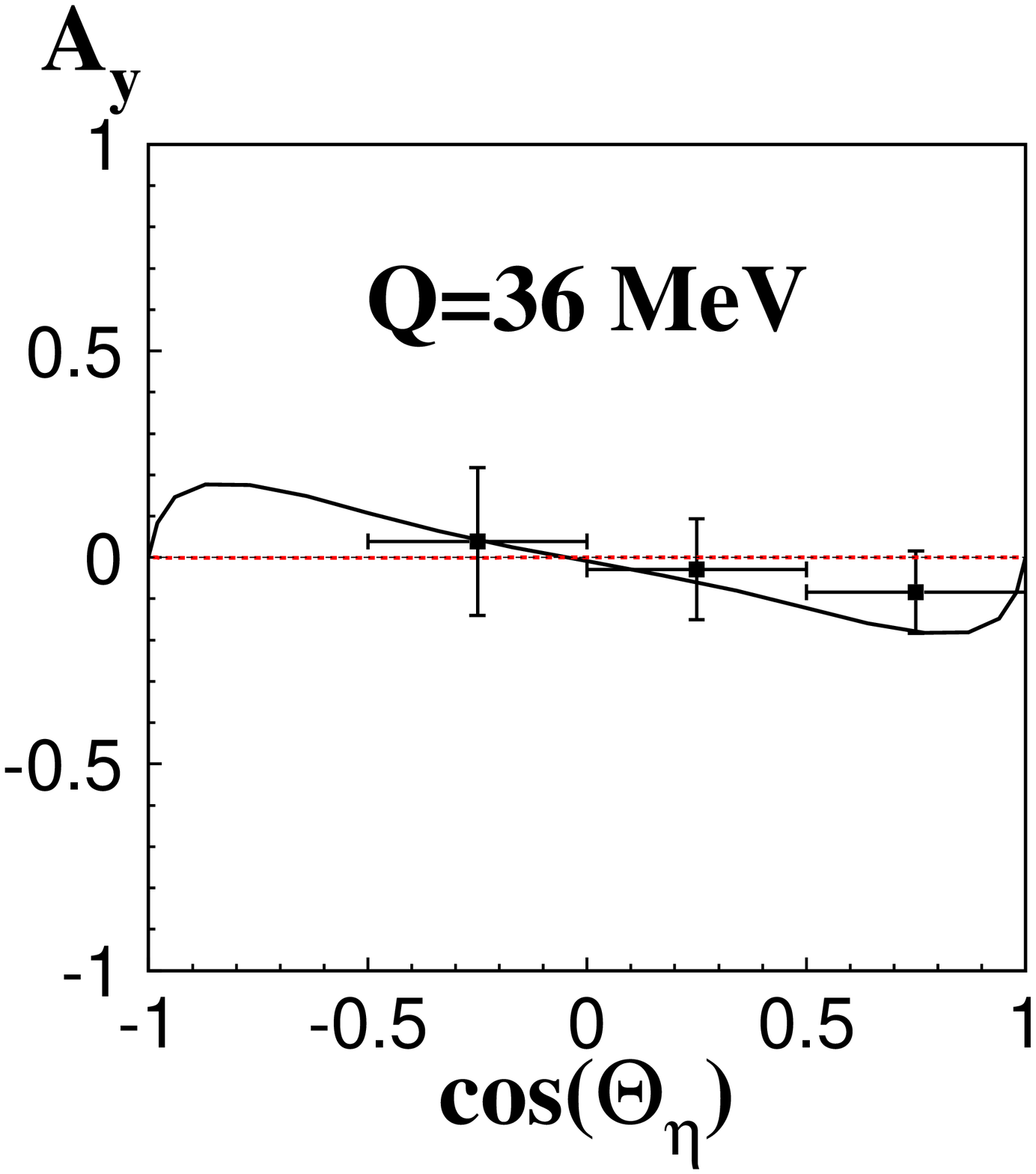}

\vspace{-0.4cm}
\caption{Analysing powers for the $\vec{p}p\to pp\eta$ reaction as functions of $\cos{\theta_{\eta}}$ for
        %the cosine of center-of-mass polar angle of the $\eta$ meson for
        Q~=~10~MeV (left panel) and Q~=~36~MeV (right panel). Full lines are the
        predictions based on the pseudoscalar meson exchange model~\cite{nakayama}
        whereas the dotted lines represent the 
        calculations based on the vector meson exchange~\cite{wilkin}.
	In the right pannel the dotted line is consistent with zero. 
        Shown are the statistical uncertainties solely. 
	%Error bars in this figure show the statistical uncertainties only.
        %Systematic errors are quoted in Table~\ref{tab1} and they are 
        %about an order of magnitude smaller than the statistical uncertainties. 
	\label{ay}}
\vspace{-0.2cm}
\end{figure}

The analysing powers calculated using Eq.~\ref{ayform}
are summarized in Table~\ref{tab1} 
for both excess energies 
and presented in Fig.~\ref{ay}.
At the excess energy of Q~=~36~MeV insufficient statistics 
for the $\cos\theta_{\eta}\in\left(-1;-0.5\right)$ 
range resulted in an error larger than the allowed range and hence
this point was omitted.

In order to verify the correctness of the models 
based on the dominance of the $\rho$ or $\pi$ meson exchanges,
a $\chi^2$ test has been performed. 
The reduced value of the $\chi^2$ for the pseudoscalar 
meson exchange model was determined to be 
$\chi^2_{psc}=0.54$, which corresponds to a 
significance level $\alpha_{psc}=0.81$, whereas 
for the vector meson exchange model $\chi^2_{vec}=2.76$, 
resulting in a significance level of $\alpha_{vec}=0.006$. 

\begin{figure}[h]
\includegraphics[width=4.27cm]{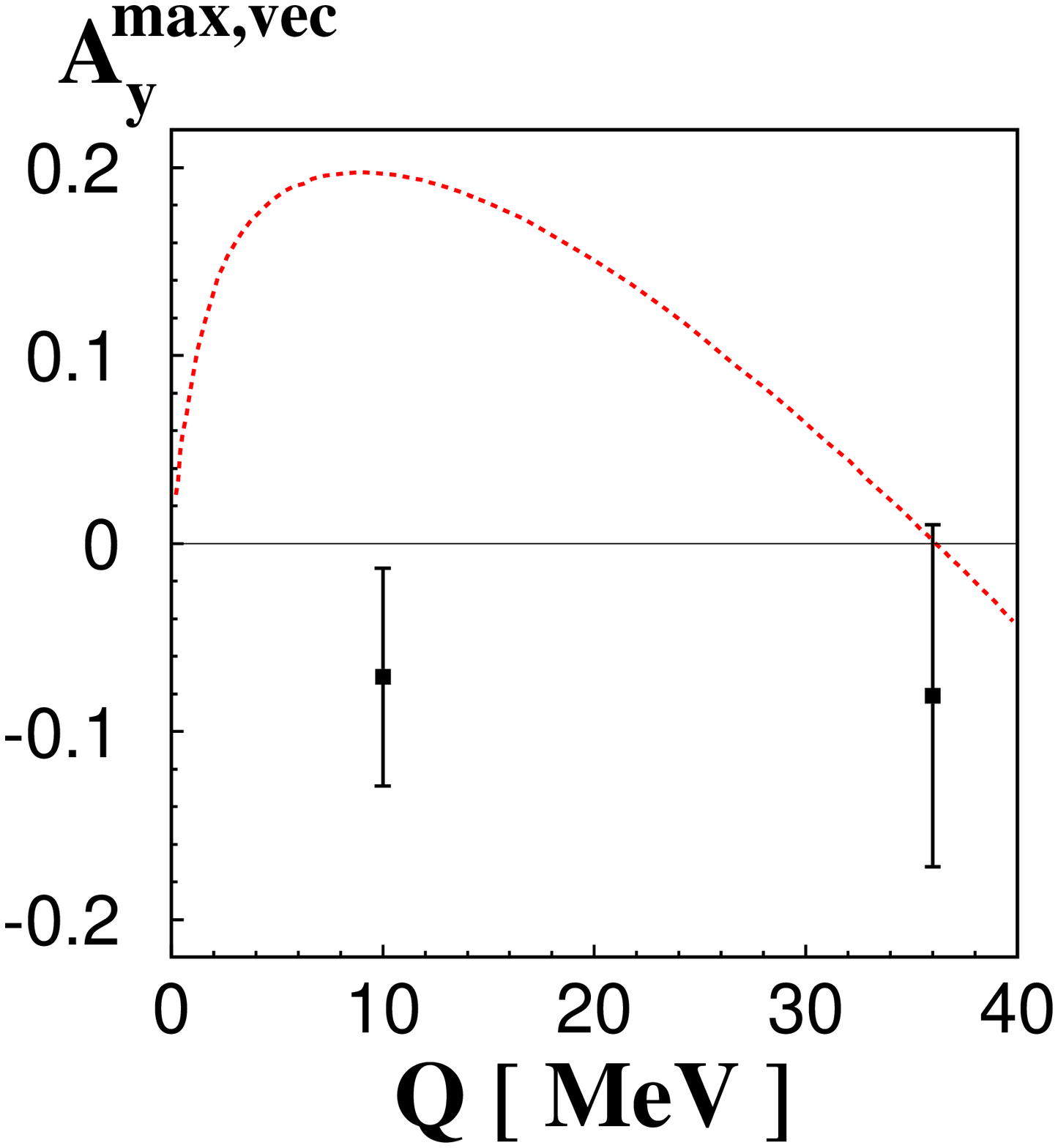}
\includegraphics[width=4.27cm]{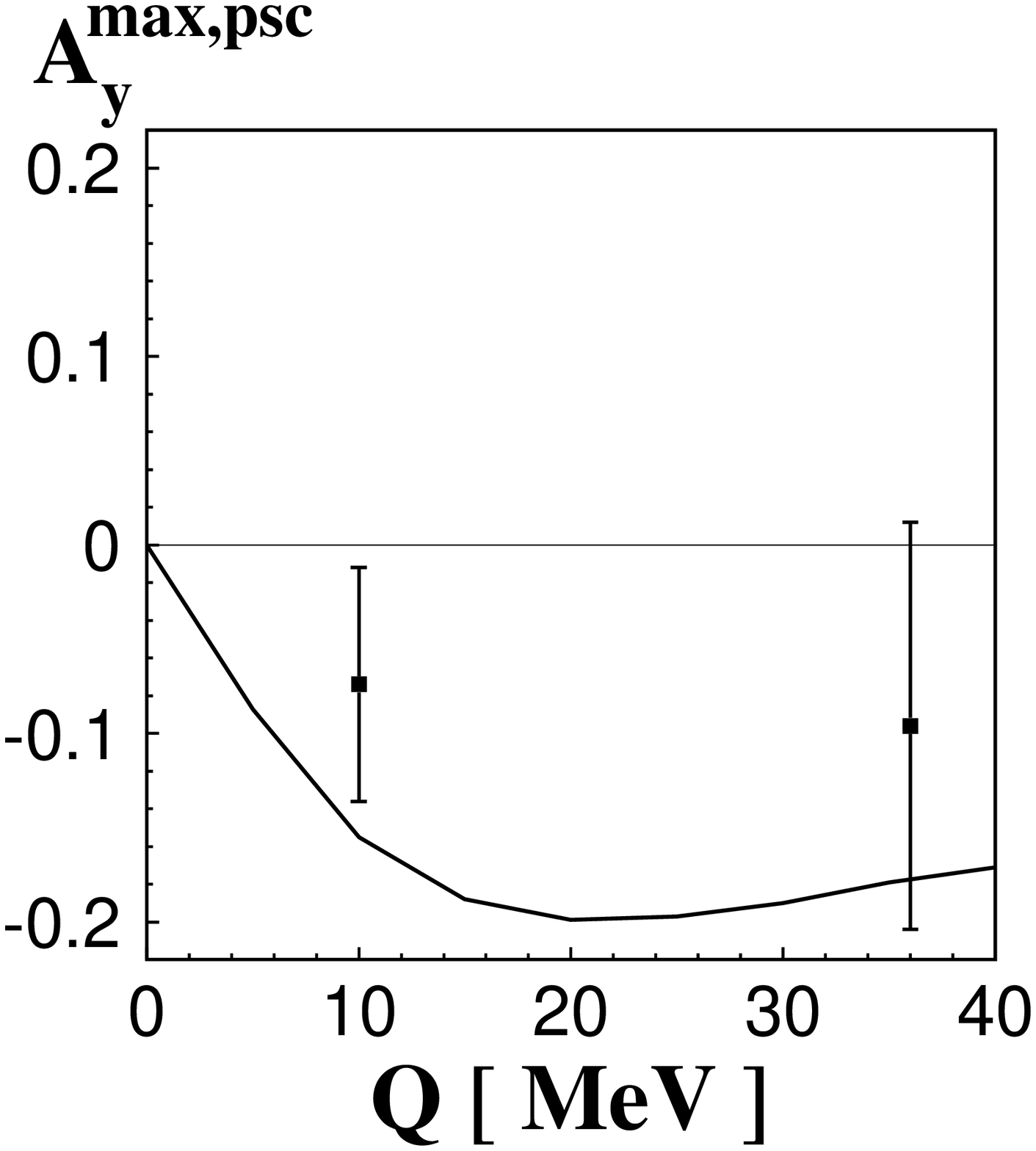}

\vspace{-0.3cm}
\caption{Theoretical predictions for the amplitudes of the analysing power's
        energy dependence 
        %in the close-to-threshold region along i
        confronted with the
        amplitudes determined in the experiments at Q~=~10 and Q~=~36~MeV
        %Full line in the right panel is the prediction of the pseudoscalar meson
        %exchange dominance model~\cite{nakayama}, while the dotted line in
        %the left panel is the prediction of the vector meson dominance model 
	%of reference~\cite{wilkin}. 
	for the vector (left panel) and pseudoscalar (right panel) 
        meson exchange dominance model. \label{ay_max}}
\end{figure}

In the vector meson exchange dominance model~\cite{wilkin} 
the angular distribution of the analysing power
is parameterized with the following equation:
\begin{equation}
A_y(\theta_{\eta}) = A_y^{max,vec} \sin 2\theta_{\eta},
\label{aymax}
\end{equation}
where the amplitude A$_y^{max,vec}$ is a function of the excess energy Q, 
shown as a dotted line in the left panel of Fig.~\ref{ay_max}.

We have estimated the values of A$_y^{max,vec}$ comparing the experimental data
with predicted shape utilizing a $\chi^2$ test. The values
of \mbox{A$_y^{max,vec}$ for Q~=~10} and 36~MeV have been determined 
to be \mbox{A$_y^{max,vec}(Q=10)~=~$-0.071~$\pm~$0.058} and  
\mbox{A$_y^{max,vec}(Q=36)~=~$-0.081~$\pm$~0.091}, respectively. 

Similar calculations have been performed for the pseudoscalar meson 
exchange model~\cite{nakayama}, assuming that the shape of the analysing power 
as a function of the $\cos\theta_{\eta}$ does not depend on the 
excess energy, which is correct within about 5\% accuracy.
It has been found that A$_y^{max,psc}(Q=10)~=~$-0.074~$\pm$~0.062,
and A$_y^{max,psc}(Q=36)~=~$-0.096$~\pm~$0.108. 
These results are presented in Fig.~\ref{ay_max}.
The figure shows
that the predictions of the model based on the $\pi$ 
mesons dominance are fairly consistent
with the data,  whereas  
the calculations based on the dominance of the $\rho$ meson exchange 
differ from the data by more than four standard deviations.
 
  Summarizing, the $\chi^2$ analysis applied to the tested production models 
  excludes the correctness of the assumption
  of a pure vector meson dominance ($\rho$ exchange)
  with a significance level of 0.006 corresponding 
  to discrepancy between the model and the data 
  larger than four standard deviations,
  and provides strong evidence for 
  the supposition
  that the production of  the $\eta$ mesons in  
  nucleon-nucleon collision is dominated by the pion exchange.
This result confirms empirically the scenario which could have been derived 
intuitively from the examination of the branching fractions of the S$_{11}$(1535) resonance
intermediating the production of the $\eta$ meson. 
In the review of particles
properties~\cite{PDG} it is quoted that S$_{11}$(1535)
decays in about 45\% to the $N\pi$ channel whereas the $\rho$N branching fraction is estimated
to be lower than 4\%.
 
One should, however, keep in mind that the interference in the exchange of both types of mesons
are not excluded and should be studied theoretically and experimentally by the measurement 
of further spin observables. 

It is also worth to mention that the analysing powers of the $\vec{p}p\to pp\eta$  reaction
for both excess energies studied are consistent with zero 
within one standard deviation. This may suggest that the $\eta$
meson is predominantly produced in the $s$-wave, an observation which  
is in agreement with the results of the analysing power measurements 
performed by the DISTO collaboration~\cite{disto} where, interestingly, 
in the far-from-threshold energy region the A$_y$ were found 
to be also consistent with zero within one standard deviation.

\begin{acknowledgments}
The authors would like to thank \mbox{K. Nakayama}, \mbox{K. S. Man},
and \mbox{C. Wilkin}
for providing the 
predictions for A$_{y}$ and A.~Kup{\'s}{\'c} and S.~Bass 
for reading the early version of the manuscript.
We acknowledge the support of the
European Community-Research Infrastructure Activity
under the FP6 programme (Hadron Physics, N4:EtaMesonNet, 
RII3-CT-2004-506078) and the support 
of the Polish State Committee for Scientific Research
(grant No. PB1060/P03/2004/26).
\end{acknowledgments}

\bibliography{czyzykiewicz}

\end{document}